\documentclass[pra,twocolumn,showpacs,showkeys,preprintnumbers,amsmath,amssymb]{revtex4}
\usepackage{tipa}
\usepackage{amssymb}
\usepackage{txfonts}
\usepackage{CJK}

\usepackage{hyperref}
\usepackage{graphicx}
\usepackage{epsfig}

\begin{document}\begin{CJK*}{GBK}{song}
\title{Quantum state transfer between three ring-connected atoms}

\author{Guo Yan-Qing$^{*}$, Deng Yao, Pei Pei\footnotetext{ $^{*}$ Corresponding author. Email:
yqguo@dlmu.edu.cn}, Wang Dian-Fu} \affiliation{Department of
Physics, Dalian Maritime University, Dalian, 116026}

\begin{abstract}
A robust quantum state transfer scheme is discussed for three
atoms that are trapped by separated cavities linked via optical
fibers in ring-connection. It is shown that, under the effective
three-atom Ising model, arbitrary quantum state can be transferred
from one atom to another deterministically via an auxiliary atom
with maximum unit fidelity. The only required operation for this
scheme is replicating turning on/off the local laser fields
applied to the atoms for two steps with time cost
$\frac{\sqrt{2}\pi}{\Gamma_{0}}$. The scheme is insensitive to
cavity leakage and atomic position due to the condition $\Delta
\approx \kappa\gg g$. Another advantage of this scheme is that the
cooperative influence of spontaneous emission and operating time
error can reduce the time cost for maximum fidelity and thus speed
up the implementation of quantum state transfer.
\end{abstract}
\maketitle

Long-range communication channels between distant qubits are
essential for practical quantum information processing. One of the
most important goal for constructing the channels is the
implementation of quantum state transfer (QST) from one qubit to
another in a deterministic way, especially for unknown quantum
state $^{[1-11]}$. Many schemes that based on spin systems that
including Heisenberg model or Ising Model, or atom-photon systems
that including cavity QED systems have been proposed to implement
QST between spins in quantum dots$^{[12-14]}$, atoms or photons in
cavities$^{[15-17]}$. The advantage of spin systems is the spins
can be easily controlled through magnetic field due to the simple
and regular interaction between neighboring spin sites. While,
differing from the short-range communication channels of spin
systems, QED systems that including intra-cavity atoms connected
via optical fibers extend QST to macroscopic length scale that is
necessary for long range quantum communications. We believe a
quantum model, such as the model suggested by Zhong et al
$^{[18]}$, combined with these two kinds of systems should be of
much importance for QST process.

However, there is a disadvantage in the schemes using QED systems,
that the schemes usually work in a probabilistic way. One of the
ways to improve the success probability and fidelity is
constructing precisely controlled coherent evolutions of the
global system and weaken the affect of probing impulse detection
inefficiency. One kind of the controlled evolutions are dominated
based on global control of the system. For example, in the scheme
considered by Serafini et al $^{[1]}$, the technic turning off the
interaction between atoms in separated cavities is used to
implement quantum swap gate and C-phase gate. In the scheme
proposed by Yin and Li$^{[9]}$, deterministic QST can be achieved
through turning off the interaction between distant atomic groups.
In the scheme proposed by Bevilacqua and Renzoni, laser pulses are
used to implement QST $^{[16]}$. Another kind of the controlled
evolutions are dominated based on local control of the system. For
example, in the scheme proposed by Mancini and Bose $^{[2]}$, the
only required control to obtain maximally entangled states is
synchronously turning on and off of the locally applied laser
fields applied on individual intro-cavity atoms.

In the present paper, we propose an alternative QST scheme based
on a simple quantum network consists of three distant atoms
trapped in distinct cavities. Such a system is treated as an
effective spin-spin interacting Ising model for distant atoms. A
two-step operation consisting of simply replicating turning on/off
of the local laser fields is put forward to implement QST between
two atoms. We demonstrate that the scheme works in deterministic
way with high fidelity. We also investigate the affect of atomic
spontaneous emission on the fidelity of the scheme.

We firstly recall the model put forward in Ref. [19]. The
schematic setup of the model is shown in Fig. 1. Three two-level
atoms 1, 2 and 3 are trapped in spatially separated optical
cavities which are assumed to be single-sided. Atoms interact with
cavity field in a dispersive way. Three off-resonant driving
external fields are added on cavities. Two neighboring cavities
are connected via optical fiber. The global system is located in
vacuum.
\begin{figure}
\epsfig{file=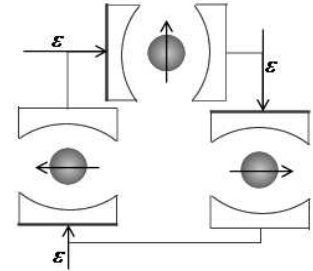, width=4cm,
height=3.7cm,bbllx=0,bblly=0,bburx=182,bbury=172}\caption{{\protect\footnotesize
{Schematic setup of the supposed system. Three two-level atoms are
trapped in separate optical cavities linked via optical fibers.
Each of the cavities is driven by an external field. Every atom is
coupled to a local laser field.}}}
\end{figure}
Using the input-output theory, taking the adiabatic approximation
$^{[20]}$ and applying the methods developed in Refs. [2], we
obtain the effective Hamiltonian of the global system as
$H_{eff}=H_{zz}$, where $H_{zz}=J(\sigma _{1}^{z}\sigma
_{2}^{z}+\sigma _{2}^{z}\sigma _{3}^{z}+\sigma _{3}^{z}\sigma
_{1}^{z})$, $\sigma_{i}^{z}$ is spin operators of atom i. And
$J=2\kappa\chi ^{2}Im\left\{ |\alpha|^2(Me^{i\phi }+\kappa
e^{i2\phi})/(M^{3}-W^{3})\right\}$, where $\kappa$ is the cavity
leaking rate, $\chi=\frac{g^2}{\Delta}$, $g$ is the coupling
strength between atom and cavity field, $\Delta$ is the detuning.
In deducing $H_{eff}$, the condition $\Delta\approx\kappa\gg g$ is
assumed, $M=i\Delta+\kappa$, $W^{3}=\kappa ^{3}e^{i3\phi}$. $\phi
_{21}$, $\phi$ is the phase delay caused by the photon
transmission along optical fiber. And
$\alpha=\varepsilon\frac{M^2+M\kappa
e^{i\phi}+\kappa^{2}e^{i2\phi}}{M^{3}-W^3}$. Such a system is
undoubtedly an Ising ring model with uniform coupling strengthes.

It has been approved that, in an isotropic Heisenberg model, the
arbitrarily perfect QST can be achieved only by applying a
magnetic field along the spin chain $^{[13]}$. Thus, we assume
local weak laser fields are applied to resonantly interact with
the atoms. Without losing of generality, we allow a simple spatial
variation of the laser fields so that the Rabi frequencies are
different for individual atoms. The effective Hamiltonian is now
written as $H_{eff}=H_{zz}+H_{x}$, where
$H_{x}=\sum\limits_{i}\Gamma_{i}\sigma_{x},\sigma_{x}=(\sigma
_{i}^{-}+\sigma _{i}^{+})$, $\sigma_{i}^{+} (\sigma_{i}^{-})$ is
raising (lowering) operator of atom $i$. This can be interpreted
as an Ising ring with electromagnetic fields applied on individual
spins in perpendicular direction. The system plays an important
role in quantum information process since two-atom entangled
stated can be generated in such system by synchronously turning
off the local laser fields $^{[21-23]}$. This paper aims to study
the QST governed by the Hamiltonian. Under the condition
$\Gamma_{i}\ll J$, the secular part of the effective Hamiltonian
can be obtained through the transformation $UH_{x}U^{-1}$,
$U=e^{-iH_{zz}t}$, as $^{[24]}$
\begin{eqnarray}
\tilde{H}=\sum\limits_{ijk}\frac{\Gamma_{i}}{2}\sigma_{i}^{x}(1-\sigma_{j}^{z}\sigma_{k}^{z})
.
\end{eqnarray}
where the subscripts $ijk$ are permutations of $1, 2, 3$. The
straight forward interpretation of this Hamiltonian is: the spin
of an atom in the Ising ring flips \textbf{if and only if} its two
neighbors have opposite spins.

The task of arbitrary unknown quantum state transfer (QST) between
two two-level systems a and b is to accomplish the implementation
$(\alpha|e\rangle_{a}+\beta|g\rangle_{a})\otimes|g\rangle_{b}\longrightarrow
|g\rangle_{a}\otimes(\alpha|e\rangle_{b}+\beta|g\rangle_{b})$
deterministically, where $\alpha$ and $\beta$ are unknown complex
number and meet the condition of normalization, and the former in
above equation is the inputting initial state while the latter is
the outputting target state.

To this end, we assume atom 1 is inputting qubit and initially in
coherent state $\alpha|e\rangle_{1}+\beta|g\rangle_{1}$, atom 2 is
outputting qubit and initially in ground state, atom 3 is an
auxiliary qubit and initially in ground state, and suppose the
local laser field applied on atom 3 is kept zero, which leads to
an unchanged state of atom 3. The secular part of the effective
Hamiltonian can be written as
$\tilde{H}=\Gamma_{1}\sigma_{1}^{x}(1-\frac{1}{2}\sigma_{2}^{z}\sigma_{3}^{z})+\Gamma_{2}\sigma_{2}^{x}(1-\frac{1}{2}\sigma_{1}^{z}\sigma_{3}^{z})$.
The evolution of the first term of initial state
$(\alpha|e\rangle_{1}+\beta|g\rangle_{1})\otimes|g\rangle_{2}\otimes|g\rangle_{3}$
is restricted within the subspace spanned by the following basis
vectors
\begin{eqnarray}
|\phi_{1}\rangle=|e\rangle_{1}|e\rangle_{2}|g\rangle_{3},
|\phi_{2}\rangle=|e\rangle_{1}|g\rangle_{2}|g\rangle_{3},
|\phi_{3}\rangle=|g\rangle_{1}|e\rangle_{2}|g\rangle_{3},
\end{eqnarray}
while the second term remains unchanged. The Hamiltonian in Eq.
(7) is now written as
\begin{eqnarray}
\tilde{H}=\left(\begin{array}{ccc}
0 & \Gamma_{2} & \Gamma_{1}\\
\Gamma_{2} & 0 & 0\\
\Gamma_{1} & 0 & 0\\
\end{array}\right).
\end{eqnarray}

The eigenvalues of the Hamiltonian can be obtained as
$E_{12}=\pm\sqrt{\Gamma_{1}^{2}+\Gamma_{2}^{2}}$,$E_{3}=0$, and
the corresponding eigenvectors are
\begin{eqnarray}
|\psi\rangle_{i}=\sum\limits_{j}S_{ij}|\phi_{j}\rangle
\end{eqnarray}
where
\begin{eqnarray}
S=\left(\begin{array}{ccc} \frac{1}{\sqrt{2}} &
\frac{\Gamma_{2}}{\sqrt{2(\Gamma_{1}^{2}+\Gamma_{2}^{2})}} &
\frac{\Gamma_{1}}{\sqrt{2(\Gamma_{1}^{2}+\Gamma_{2}^{2})}}
\\
-\frac{1}{\sqrt{2}} &
\frac{\Gamma_{2}}{\sqrt{2(B_{1}^{2}+\Gamma_{2}^{2})}} &
\frac{\Gamma_{1}}{\sqrt{2(\Gamma_{1}^{2}+\Gamma_{2}^{2})}}
\\
0 & -\frac{\Gamma_{1}}{\sqrt{\Gamma_{1}^{2}+\Gamma_{2}^{2}}} &
\frac{\Gamma_{2}}{\sqrt{\Gamma_{1}^{2}+\Gamma_{2}^{2}}} \\
\end{array}\right)
\end{eqnarray},
which represents unitary transformation matrix between
eigenvectors and basis vectors. For initial system state
$|\Psi(0)\rangle=\sum\limits_{i}c_{i}(0)|\phi_{i}\rangle$, the
evolving global system state can be written as
$|\Psi(t)\rangle=\sum\limits_{i}c_{i}(t)|\phi_{i}\rangle$ and is
governed by the Schr\"{o}dinger equation $i\frac{\partial
|\Psi(t)\rangle}{\partial t}=\tilde{H}|\Psi(t)\rangle$. The
coefficients $c_{i}(t)$ are then given by $^{[9]}$
$c_{i}(t)=\sum\limits_{j}[S^{-1}]_{ij}[Sc(0)]_{j}e^{-iE_{j}t}$,
where $c(0)=[c_{1}(0),c_{2}(0),c_{3}(0)]^{T}$.

For initial coefficients $c(0)=[0,\alpha,0]^{T}$, the coefficients
can be obtained as
\begin{eqnarray}
c_{1}(t)&=&-i\frac{\alpha\Gamma_{2}}{\Omega}\textrm{sin}\Omega t,\nonumber\\
c_{2}(t)&=&\frac{\alpha\Gamma_{1}^{2}}{\Omega^{2}}+\frac{\alpha\Gamma_{2}^{2}}{\Omega^{2}}\textrm{cos}\Omega t\nonumber \\
c_{3}(t)&=&\frac{-\alpha\Gamma_{1}\Gamma_{2}}{\Omega^{2}}+\frac{\alpha\Gamma_{1}\Gamma_{2}}{\Omega^{2}}\textrm{cos}\Omega
t.
\end{eqnarray}
where $\Omega=\sqrt{\Gamma_{1}^{2}+\Gamma_{2}^{2}}$.

It is easily shown that, one can take
$\Gamma_{1}=\Gamma_{2}=\Gamma_{0}$ and turn off the local laser
fields applied to atom 1 and atom 2 synchronously at
$t_{p}=\frac{(2k-1)\pi}{\Omega}, k=1,2,3, ...$ and obtain the
system state as
\begin{eqnarray}
|\Psi(t_{p_{0}})\rangle=|g\rangle_{1}\otimes(-\alpha|e\rangle_{2}+\beta|g\rangle_{2})|g\rangle_{3}.
\end{eqnarray}

The above state differs from the target state
$|g\rangle_{1}\otimes(\alpha|e\rangle_{2}+\beta|g\rangle_{2})\otimes|g\rangle_{3}$
due to a minus sign. To obtain the target state exactly, one may
program the operating process as in Table. 1 (the term '$\pi$
pulse' in the table is used to denote an equivalent evaluating
time $t_{p_{0}}=\frac{\pi}{\Omega}$):

\begin{tabular}{l|l}
\multicolumn{2}{l}{Table. 1 Operation sequence for implementing QST}\\
\hline
operation sequence & system state\\
\hline $\Gamma_{1}=\Gamma_{3}=\Gamma_{0},\Gamma_{2}=0$ & initial
state
$(\alpha|e\rangle_{1}+\beta|g\rangle_{1})|g\rangle_{2}|g\rangle_{3}$\\
\hline $\pi$ pulse on atoms 1 and 3 &
$|g\rangle_{1}|g\rangle_{2}(-\alpha|e\rangle_{3}+\beta|g\rangle_{3})$\\
\hline $\Gamma_{2}=\Gamma_{3}=\Gamma_{0},\Gamma_{2}=0$ &
$|g\rangle_{1}|g\rangle_{2}(-\alpha|e\rangle_{3}+\beta|g\rangle_{3})$\\
\hline $\pi$ pulse on atoms 2 and 3 &
target state $|g\rangle_{1}(\alpha|e\rangle_{2}+\beta|g\rangle_{2})|g\rangle_{3}$\\
\hline
\end{tabular}

The above operating process can be interpreted as two steps:

Firstly, turning on the laser field acting on atom 1 and 3 while
keeping the laser field acting atom 3 zero. At the specific time
$t_{p_{1}}=\frac{(2k-1)\pi}{\Omega}, k=1,2,3, ...$, turning off
the laser fields synchronously.

Secondly, turning on the laser field acting on atom 2 and 3 while
keeping the laser field acting atom 1 zero. At the specific time
$t_{p_{2}}=\frac{(2k-1)\pi}{\Omega}, k=1,2,3, ...$, turning off
the laser fields synchronously.

In this procedure, one do not need to know the values of
coefficients $\alpha$ and $\beta$, and do not require any methods
of quantum coincidence measurement on atoms. An unknown QST is
implemented deterministically with $100\%$ success probability.

To illustrate the the efficiency of the QST, not only at specific
times, but also in overall view of time scales, we plot the
fidelity of QST between atoms 1 and 2 with respect to operating
times $t_{1}$, which represents the operating time of the first
step, and $t_{2}$, which denotes that of the second step. The
average fidelity is defined as $^{[9]}$
\begin{eqnarray}
F=\frac{1}{2\pi}\int_{0}^{2\pi}|\langle\Psi_{f}|\Psi(t)\rangle|^{2}
d\theta
\end{eqnarray}
where, $|\Psi_{f}\rangle$ is the target state. In Fig. 2 (a), it
can be seen that the overall quantity of fidelity is governed by
operating times $t_{p_{1}}$ and $t_{p_{2}}$ almost equally. The
fidelity periodically reaches the maximum 1 at specific times
$t_{p_{1}}=t_{p_{2}}=\frac{\pi}{\Omega}$. The time cost of the
scheme can be estimated as $t_{p}\approx \frac{2\pi}{\Omega}$.

The above results explicitly demonstrate a deterministic two-step
QST scheme between atoms 1 and 2 which is accomplished by only
turning on two identical local laser fields applied on atoms 1 and
3 and turning them off at typical times synchronously, and
duplicate the step for atoms 2 and 3. Similarly, QST between atoms
2 and 3 or between atoms 1 and 3 can be accomplished similarly.

So, the total procedure of the scheme implementation consists only
two steps: a step turning on/off the laser fields synchronously
for inputting atom and auxiliary atom and repeat the step for
auxiliary atom and outputting atom.
\begin{figure}
\epsfig{file=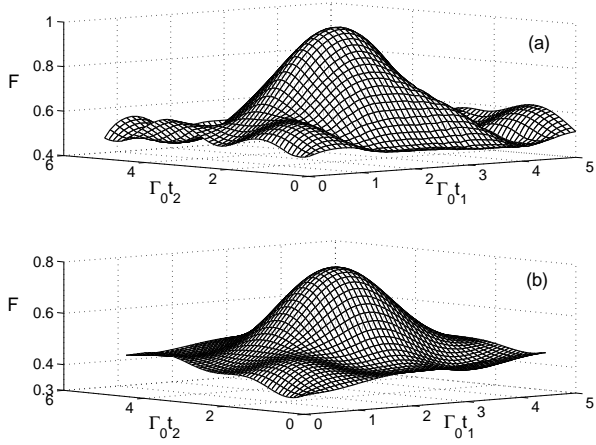, width=8cm,
height=6cm,bbllx=35,bblly=202,bburx=560,bbury=588}\caption{{\protect\footnotesize
{Fidelity of QST between atoms 1 and 2 with respect to operating
times $t_{1}$ and $t_{2}$ for (a) atomic spontaneous emission rate
$\gamma=0$, (b) atomic spontaneous emission rate
$\gamma=0.1\Gamma_{0}$}}}
\end{figure}

In this model, the leakage of cavity fields is assumed to be large
enough to keep the validity of the adiabatic approximation for
obtaining effective Hamiltonian. While, the inevitable atomic
spontaneous emission still challenges the efficiency of the scheme
and results in a dissipative effect, which can be estimated by
adding a non-Hermitian conditional term to the Hamiltonian in Eqn.
(1)$^{[25]}$. The global Hamiltonian can be written as
$H_{s}=-i\gamma\sum\limits_{i}|e\rangle_{i}\langle e|+\tilde{H}$,
where $\gamma$ represents the atomic spontaneous emission rate. In
the subspace spanned by
$|\phi_{1}\rangle=|e\rangle_{1}|g\rangle_{2}|e\rangle_{3}$,
$|\phi_{2}\rangle=|e\rangle_{1}|g\rangle_{2}|g\rangle_{3}$,
$|\phi_{3}\rangle=|g\rangle_{1}|g\rangle_{2}|e\rangle_{3}$, for
initial state $|e\rangle_{1}|g\rangle_{2}|g\rangle_{3}$, the
evolved coefficients can be obtained as
\begin{eqnarray}
c_{1}(t)&=&-\frac{i\alpha\Gamma_{3}}{\Lambda}e^{-\frac{3}{2}\gamma
t}\textrm{sin}\Lambda t,\nonumber \\
c_{2}(t)&=&\frac{\alpha\Gamma_{1}^{2}}{\Omega^{2}}e^{-\gamma
t}(1+\frac{\Gamma_{3}^{2}}{\Gamma_{1}^{2}}e^{-\frac{\gamma
t}{2}}\textrm{cos}\Lambda t
+\frac{\gamma}{\Lambda}\frac{\Gamma_{3}^{2}}{\Gamma_{1}^{2}}e^{-\frac{\gamma
t}{2}}\textrm{sin}\Lambda t),\nonumber \\
c_{3}(t)&=&\frac{\alpha\Gamma_{1}\Gamma_{3}}{\Omega^{2}}e^{-\gamma
t}(-1+e^{-\frac{\gamma t}{2}}\textrm{cos}\Lambda t
+\frac{\gamma}{\Lambda}e^{-\frac{\gamma t}{2}}\textrm{sin}\Lambda
t),
\end{eqnarray}
where
$\Lambda=\sqrt{\Gamma_{1}^{2}+\Gamma_{3}^{2}-\frac{\gamma^{2}}{4}}$.

Taking $\Gamma_{1}=\Gamma_{3}=\Gamma_{0}$ and shutting down the
laser fields applied to atom 1 and atom 3 synchronously at
specific time $t_{p_{1}}=\frac{(2k-1)\pi}{\Lambda}, k=1,2,3, ...$
, one can obtain the system state as
\begin{eqnarray}\Psi(t_{p_{1}})&=& \alpha A_{1}e^{-\gamma
t_{p_{1}}}|e\rangle_{1}|g\rangle_{2}|g\rangle_{3}\nonumber\\
&-&\alpha B_{1}e^{-\gamma
t_{p_{1}}}|g\rangle_{1}|g\rangle_{2}|e\rangle_{3}+\eta\beta
|g\rangle_{1}|g\rangle_{2}|g\rangle_{3},
\end{eqnarray}
where $\eta$ is an additional normalized factor,
$A_{1}=\frac{(1-e^{-\frac{\gamma t_{p_{1}}}{2}})}{2},
B_{1}=\frac{(1+e^{-\frac{\gamma t_{p_{1}}}{2}})}{2}$. Now, we let
the above state be new inputting initial state without delay and
take $\Gamma_{2}=\Gamma_{3}=\Gamma_{0}$, and shutting down the
laser fields applied to atom 2 and atom 3 synchronously at
specific time $t_{p_{2}}=\frac{(2k-1)\pi}{\Lambda}, k=1,2,3, ...$.
It can be proved that, under this condition, there is no
transition between the first term in Eqn. (10) and other
three-atom excited states such as
$|e\rangle_{1}|e\rangle_{2}|g\rangle_{3}$,
$|e\rangle_{1}|g\rangle_{2}|e\rangle_{3}$. After some complicated
calculation, the system state can be obtained analytically as
\begin{eqnarray}
|\Psi(t_{p})&=&\alpha
e^{-\gamma(t_{p_{1}}+t_{p_{2}})}B_{1}B_{2}|g\rangle_{1}|e\rangle_{2}|g\rangle_{3}\\
\nonumber &-&\alpha
e^{-\gamma(t_{p_{1}}+t_{p_{2}})}B_{1}A_{2}|g\rangle_{1}|g\rangle_{2}|e\rangle_{3}\\
\nonumber &-&\alpha
e^{-\gamma(t_{p_{1}}+3t_{p_{2}})}A_{1}|e\rangle_{1}|g\rangle_{2}|g\rangle_{3}+\lambda
\beta |g\rangle_{1}|g\rangle_{2}|g\rangle_{3},
\end{eqnarray}
where $A_{2}=\frac{(1-e^{-\frac{\gamma t_{p_{2}}}{2}})}{2},
B_{2}=\frac{(1+e^{-\frac{\gamma t_{p_{2}}}{2}})}{2}$, $\lambda$ is
an additional normalized factor. The time cost of the scheme is
now estimated as $t_{p}=t_{p_{1}}+t_{p_{2}}\approx
\frac{2\pi}{\Lambda}$.

\begin{figure}
\epsfig{file=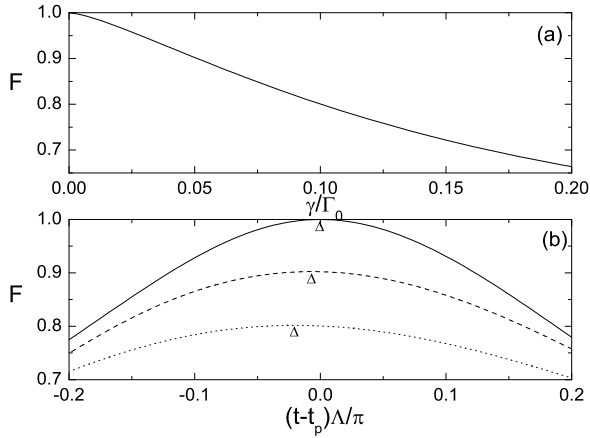, width=8cm,
height=6cm,bbllx=10,bblly=14,bburx=305,bbury=240}\caption{{\protect\footnotesize
{Fidelity of QST between atoms 1 and 2 with respect to (a) atomic
spontaneous emission, the operating time
$t_{p_{1}}=t_{p_{2}}=\frac{\pi}{\Lambda}$, and (b) operating time
error $(t-t_{p})\Lambda/ \pi$, where the triangle indicates the
maximum value of fidelity, $\gamma=0$ for solid line,
$\gamma=0.05\Gamma_{0}$ for dashed line, $\gamma=0.1\Gamma_{0}$
for dotted line.}}}
\end{figure}

In Fig. 2 (b), we plot the fidelity of QST with respect to
operating time $t_{p_{1}}$ and  $t_{p_{2}}$ for atomic spontaneous
emission rate $\gamma=0.1\Gamma_{0}$. Obviously, the atomic
spontaneous emission reduces the overall quantity of fidelity and
smoothing the oscillation of the fidelity. In Fig. 3 (a), we plot
fidelity of QST between atoms 1 and 2 with respect to atomic
spontaneous emission. Obviously, the spontaneous emission
monotonically decreases the maximum quantity of fidelity of QST,
which corresponds to the strict operating time condition
$t_{p_{1}}=t_{p_{2}}=\frac{\pi}{\Lambda}$.  While, in practical
case, operating time error emerges inevitably. The existence of
operating time error can lead to a more complicated system state
includes $|e\rangle_{1}|e\rangle_{2}|g\rangle_{3}$,
$|e\rangle_{1}|g\rangle_{2}|e\rangle_{3}$,
$|g\rangle_{1}|e\rangle_{2}|e\rangle_{3}$,
$|e\rangle_{1}|g\rangle_{2}|g\rangle_{3}$,
$|g\rangle_{1}|g\rangle_{2}|e\rangle_{3}$,
$|g\rangle_{1}|e\rangle_{2}|g\rangle_{3}$. Only the last term
contributes to the fidelity of QST. In Fig. 3 (b), we plot
fidelity of QST between atoms 1 and 2 with respect to operating
time error $(t-t_{p})\Lambda/ \pi$ for different atomic
spontaneous emission rates. It is interesting that operating time
error, for larger spontaneous emission, on one hand decreases the
maximum quantity of fidelity, on the other hand reduces the time
cost for achieving maximum quantity of fidelity and, in other
words, speeds up the implementation of QST, which is the
cooperative influence of spontaneous emission and operating time
error. Further more, the sensitivity of fidelity to operating time
error is decreased for larger spontaneous emission.

In summary, we have discussed an arbitrary QST scheme in a system
contains three distant atoms by simply replicating the operation
of synchronously turning on/off the locally applied laser fields
for individual atoms. The auxiliary atom is used to avoid
additional single qubit phase shift operation and the resulting
QST is deterministic and in 100\% fidelity. We discuss the affect
of atomic spontaneous emission on QST. It is shown that the atomic
spontaneous emission decreases the quantity of fidelity, while the
cooperative influence of spontaneous emission and operating time
error reduces the time cost $\frac{2\pi}{\Lambda}$ for maximum
fidelity and thus speeds up the implementation of QST. It has been
demonstrated that the dissipation of the photon leakage along
optical fibers can be included in the spin-spin coupling
coefficients by replacing the phase factor $e^{i\phi}$ in Eq. (3)
with $e^{i\phi-\nu L}$ $^{[2]}$ , where $\nu$ is the fiber decay
per meter and $L$ is the length of the fiber between atoms $i$ and
$j$. For typical fibers $^{[26]}$, the decay per meter is
$\nu\thickapprox 0.08$. The spin-spin coupling coefficient
$J_{0}^{'}$  is now about $90\%$ of $J_{0}$. The rotating wave
approximation in deriving secular part of effective Hamiltonian is
still kept valid under the condition $\Gamma_{i} \ll
J_{0}^{'}\thickapprox 0.9 J_{0}$. So the QST gate still works with
high fidelity. Furthermore, we have assumed $\kappa \gg g$ in the
calculation of deriving effective Ising model, which ensures the
scheme is insensitive to the slight variation of strong leakage
rate. As is concluded, the scheme works in a robust way since both
the affected aspects of fiber lossy and cavity dissipation can be
neglected. It should also be noticed that to avoid the inevitable
time-delay affect caused by mismatch of practical and theoretical
controlling times $^{[27]}$, remedial methods such as Lyapunov
control can be used in the extended scheme. Many of the present
schemes only contain two atoms trapped in separated cavities. From
a realistic point of view, a robust quantum network must contains
many distant quantum nodes. QST must be implemented between any
two quantum nodes in high fidelity. The model used and the results
obtained in this scheme may act as a possible candidate.

We thank Professor Dian Min Tong and Professor Dong Mi for helpful
discussions and their encouragement. This work is supported by NSF
of China under Grant No. 11305021 and the Fundamental Research
Funds for the Central Universities.

\end{CJK*}
\end{document}